\documentclass[8pts,superscriptaddress,aps,pra,twocolumn,showpacs,nofootinbib,longbibliography]{revtex4-1}
 \usepackage{etex}
 \usepackage{amsmath,amssymb,amsthm}
 \usepackage[colorlinks=true,citecolor=blue,urlcolor=blue]{hyperref}
 \usepackage[pdftex]{graphicx}
 \usepackage{times,txfonts}
 \usepackage{braket}
 \usepackage{color}
 \usepackage{graphics,epstopdf}
 \usepackage{natbib}
 \usepackage{pgfplots}
 \usepackage{tikz-3dplot}
 \usepackage{titlesec}

 \titleformat{\subsubsection}[runin]
   {\normalfont\normalsize\itshape}{\thesubsubsection}{1em}{}
 
  \tdplotsetmaincoords{60}{115}
 \pgfplotsset{compat=newest}

 \newtheorem{thm}{Theorem}
 \newtheorem{result}{Result}

\begin{document}
 
 \title{Sufficient conditions for quantum advantage in random access code protocols with two-qubit states}
 
 \author{Som Kanjilal}
 \email{som.kanjilal1011991@gmail.com}
 \affiliation{Center for Astroparticle Physics and Space Science (CAPSS), Bose Institute, Kolkata-700 091, India}
\affiliation{Harish-Chandra Research Institute,  A CI of Homi Bhabha National Institute, Chhatnag Road, Jhunsi, Prayagraj - 211019}
\affiliation{ Center for Quantum Science and Technology, International Institute of Information Technology, Gachibowli, Hyderabad 500032, India}
 
 \author{Chellasamy Jebarathinam}
 \affiliation{S. N. Bose National Centre for Basic Sciences, Block JD, Sector III, Salt Lake, Kolkata 700 106, India}
 \affiliation{Department of Physics and Center for Quantum Frontiers of Research and Technology (QFort), National Cheng Kung University, Tainan 701, Taiwan}
 \affiliation{ Center for Theoretical Physics, Polish Academy of Sciences, Aleja Lotnikow 32/46, 02-668 Warsaw, Poland}
 \affiliation{Department of Physics and Center for Quantum Information Science, National Cheng Kung University, Tainan 701, Taiwan}
 
 \author{Tomasz Paterek}
 \affiliation{School of Physical and Mathematical Sciences, Nanyang Technological University, 637371 Singapore, Singapore}
\affiliation{MajuLab, International Joint Research Unit UMI 3654, CNRS, Universite Cote d'Azur, Sorbonne Universite, National University of Singapore, Nanyang Technological University, Singapore}
\affiliation{Institute of Theoretical Physics and Astrophysics, Faculty of Mathematics, Physics and Informatics, University of Gda\'nsk, 80-308 Gda\'nsk, Poland}
\affiliation{Department of Physics, School of Mathematics and Physics, Xiamen University Malaysia, 43900 Sepang, Malaysia}

\author{Dipankar Home}
 \affiliation{Center for Astroparticle Physics and Space Science (CAPSS), Bose Institute, Kolkata-700 091, India}
 
 \begin{abstract}
Random access code (RAC) is an important communication protocol to obtain information about a randomly specified substring of an $n$-bit string, while only having limited information about the $n$-bit string.  Quantum RACs usually utilise either communication of quantum bits or a shared-in-advance quantum state used in conjunction with classical communication. Here we consider the latter version of the quantum protocols under the constraint of single-bit communication and with shared \emph{arbitrary} state of two qubits. Taking the worst-case success probability as the figure of merit, we demonstrate that any state with invertible correlation matrix can be used to outperform the best classical RAC for $n=3$.
We derive an additional condition sufficient to beat the best classical performance in the case of $n=2$. 
In particular, separable states turn out to be a useful resource behind the quantum advantage for $n=2,3$. For $n \geq 4$ RACs assisted with a single copy of a quantum state do not outperform the classical RACs.
\end{abstract}
 
 \maketitle

\section{Introduction}

An $n\xrightarrow{p}m$ RAC is a two-party communication protocol in which a sender-encoder (Alice) is supplied with a bit-string of length $n$, 
and the receiver-decoder (Bob) is asked to give the value of a randomly selected bit of Alice, upon receiving from her $m < n$ classical bits of communication. Bob's answers are required to be correct with probability at least $p$, i.e. $p$ is the worst-case success probability of the RAC protocol~\cite{W83,AA+08}. Quantum RACs utilise quantum bits to outperform the best classical protocols and have been adapted for multifaceted applications ranging from random number generation \cite{LY+11}, network coding theory \cite{HI+06}, quantum key distribution \cite{PB11} to dimension witnessing \cite{WC+08} and self-testing \cite{TK+18}. From the foundational perspective, a version of RAC in the context of generalized probability theory \cite{SW08}, studied for the Popescu-Rohrlich (PR) box~\cite{GH+14}, facilitates partial characterization of quantum theory from information theoretic principles~\cite{PP+09,p19}.

The quantum protocols usually come in two flavours (but see also e.g.~\cite{Tavakoli21}).They either utilize communication of quantum particles or shared quantum states and classical communication. In the former case, and focusing on $m=1$, only quantum protocols with $n = 2$ and $3$ outperform the classical solutions~\cite{AA+08}, whereas for $n \geq 4$ there are no strategies that would allow reading any bit of Alice with probability better than just a random guess~\cite{HIN+06}. In the latter case, many copies of maximally entangled states were shown via concatenation procedure to enable performance better than classical in any $n\xrightarrow{p}1$ code~\cite{PZ10}.

Here we also focus on $n\xrightarrow{p}1$ quantum protocols and allow sharing in advance a single copy of an arbitrary two-qubit state, not necessarily an entangled one~\cite{BP14}. Our figure of merit is the worst-case success probability. These protocols are compared against the best classical ones where a pair of classical bits is shared in advance, i.e., both assisting resources are \textit{dimensionally equivalent}. In Ref.~\cite{BP14}, under the heading \textit{finite shared randomness assisted random access codes}, a comparison of these two protocols was done. It was demonstrated that Bell diagonal states with nonzero quantum discord can be advantageous over their equivalent classical counterparts by building explicit quantum codes. However, it is unclear whether quantum discord is sufficient to guarantee the existence of efficient $n\xrightarrow{p}1$ quantum codes. Accordingly, here we enquire about characteristics of assisting arbitrary bipartite qubit states that ensure the presence of RACs outperforming the most efficient dimensionally equivalent classical protocols.

Our main result shows that all assisting states with invertible correlation matrix can be exploited for quantum advantage when $n = 3$, and we derive an additional condition sufficient for quantum advantage when $n = 2$. In both cases ($n=2,3$) we provide explicit examples of efficient quantum codes when a general bipartite qubit state with invertible correlation matrix is shared between the parties. It is found that not all discorded states are useful for the provided codes.
Since the scenario we study here can be reduced to communication of quantum systems, Ref.~\cite{HIN+06} implies that for $n \ge 4$ there are no non-trivial RACs, independently of the assisting state.  We provide an alternative proof for this impossibility. 


\section{Finite Shared Randomness Assisted Random Access Codes}
\label{sec2}

\subsection{Scenario and Basic Features}
\begin{figure}[!t]
     \includegraphics[scale=0.10]{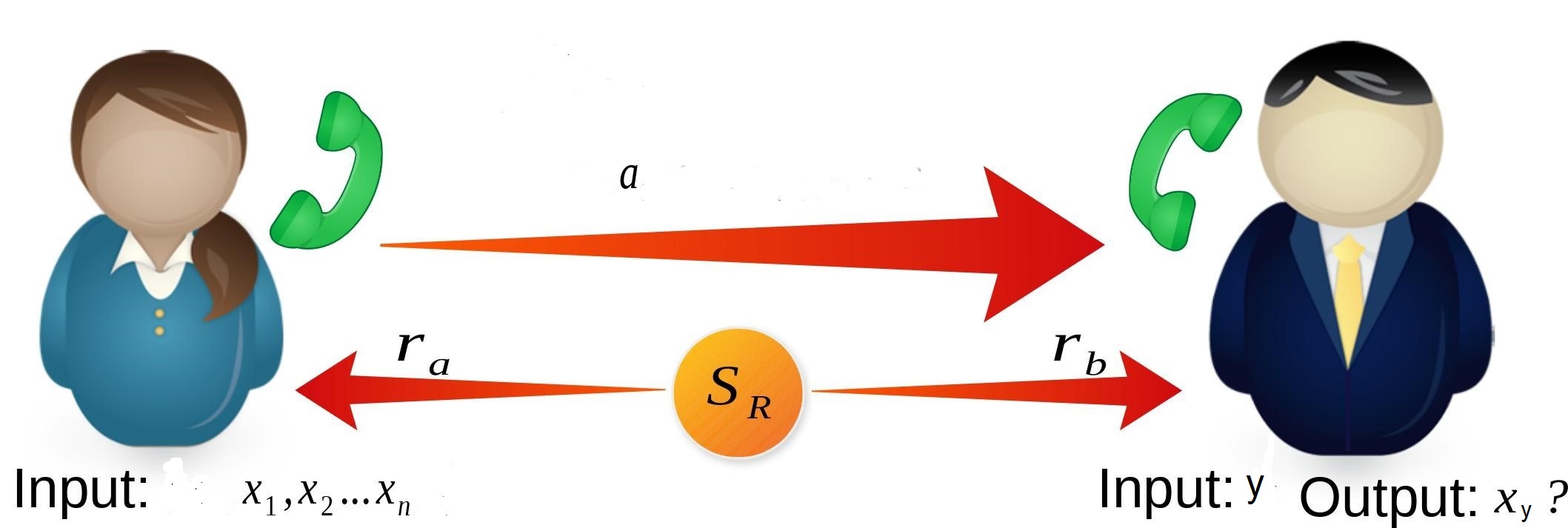}
     \caption{Shared randomness assisted $n\xrightarrow{p}1$ RAC. 
     Alice is given a dataset $x = (x_1,\dots, x_n)$ composed of $n$ classical bits $x_i$.
     Bob is asked to give the value of randomly selected bit of Alice, i.e. his input is $y$ and his guess $b$ is correct when $b = x_y$.
     To help in the process Alice communicates a single classical bit $a$ to Bob, and they are allowed to share in advance a finite amount of correlated classical or quantum bits $r_a$ and $r_b$ from a common source $S_R$.
     \label{FIG_RAC}}
\end{figure}

The scenario we study is depicted in Fig.~\ref{FIG_RAC}. Before the protocol begins Alice and Bob receive classical or quantum bits (qubits) from a common source $S_{R}$. We denote these systems as $r_{a}$ and $r_{b}$, respectively. Let $\mathcal{X}:=\{1,2,\dots,2^{n}\}$ be the set of all possible $n$ bit strings. The game starts with Alice reading her dataset of $n$ classical bits $x=[x_{1},x_{2},\ldots,x_{n}]\in \mathcal{X}$ where $x_{i}$ is either $0$ or $1$.  She encodes the information about $x$ in a single classical bit $a$, possibly using measurement results obtained by measuring the (qu)bit from the shared source. She then communicates the classical bit $a$ to Bob. Using the communicated bit and measuring his (qu)bit, previously obtained from the shared source, Bob guesses a randomly chosen bit of Alice's dataset. The protocol is successful if Bob's guess is correct.

\subsubsection{\underline{Classical version}:}
\label{cv}
 The classical version of this game, where single classical bits $r_{a}$ and $r_{b}$ are communicated from the common source to Alice and Bob respectively, has been studied in Ref.~\cite{BP14}. It was found there that the maximal classical worst-case success probabilities are the following:
\begin{itemize}
\item[(i)] $p_{\mathrm{cl}} = \frac{1}{2}$ if $n \geq 3$;
\item[(ii)]  $p_{\mathrm{cl}} = \frac{2}{3}$ if $n = 2$;
\item[(iii)] $p_{\mathrm{cl}} = \frac{1}{2}$ if $n\geq 2$ and the shared bits have maximally mixed marginals for each party, i.e., for $i=a,b$, $\text{probability of $r_{i}$ being $0$} = \text{probability of $r_{i}$ being $1$}=\frac{1}{2}$.
\end{itemize}
A brief outline of the proof for these classical upper bounds is given in Appendix~\ref{app0}.

\subsubsection{\underline{Quantum version}:}
\label{qv}
In the quantum version of this game the systems $r_{a}$ and $r_{b}$ are qubits. Depending on $x$, Alice chooses a measurement setting (Bloch vector) $\hat{A}_{x}$ to be measured on her shared qubit. She then communicates the measurement outcome $a = 0,1$ to Bob. Let us define $\mathcal{Y}:=\{1,2,\ldots,n\}$. Bob's aim is to provide the value of the randomly chosen $y$th bit of the string $x$ where $ y\in \mathcal{Y}$. In other words, Bob has to guess the value of $x_y$. To this aim Bob measures his part of the bipartite qubit state along the measurement direction $\hat{B}_{y}$ and obtains the outcome $b$. He is now in the possession of bits $a$ and $b$ as well as the input $y$. Bob constructs his guess by outputting a function $g_{y}(a,b)$~\cite{PZ10,AA+08,GH+14}. 
Bob's guess of the $y$th bit is successful if $g_{y}(a,b)=x_{y}$. 
The worst-case success probability of this protocol is given by
\begin{align}
\label{minsucprob}
p & = \min_{x\in\mathcal{X},y\in\mathcal{Y}}\left\lbrace \underset{a,b}{\sum}p\left(g_{y}(a,b)=x_{y}|x,y\right)\right\rbrace\\
\label{minisuccproba}
& = \min_{x\in\mathcal{X},y\in\mathcal{Y}}\left\lbrace \underset{a,b}{\sum}\delta_{g_{y}(a,b),x_{y}}p\left(a,b|x,y\right)\right\rbrace\\
\label{minsuccprob}
& =\min_{x\in\mathcal{X},y\in\mathcal{Y}}\left\lbrace\underset{a,b}{\sum}\delta_{g_{y}(a,b),x_{y}}\text{Tr}\left(\hat{A}_{a|x}\otimes\hat{B}_{b|y}\rho_{AB}\right)\right\rbrace,
\end{align}
where $\text{Tr}(.)$ is the trace function, $\hat{A}_{a|x}=\frac{1}{2}(\mathbb{\hat{I}}+(-1)^{a}\hat{A}_{x}.\hat{\sigma})$ is the projection operator corresponding to outcome $a$ of the setting $\hat{A}_{x}$ and $\hat{B}_{b|y}=\frac{1}{2}(\mathbb{\hat{I}}+(-1)^{b}\hat{B}_{y}.\hat{\sigma})$ is the projector pertaining to outcome $b$ of measurement setting $\hat{B}_{y}$  and  $\hat \sigma = (\sigma_x, \sigma_y, \sigma_z)$ is the vector of Pauli operators. 
There are $2^{n}$ different measurement directions $\hat{A}_{x}$ for encoding and $n$ measurement directions $\hat{B}_{y}$ for decoding. The measurements are conducted on a general two-qubit shared state which we parameterise as
\begin{equation}
 \label{EQ_2Q}
 \rho_{AB}=\frac{1}{4}[\mathbb{\hat{I}}\otimes\mathbb{\hat{I}}+\vec{M}.\hat{\sigma}\otimes\mathbb{\hat{I}}+\mathbb{\hat{I}}\otimes\vec{N}.\hat{\sigma}+\sum_{i,j}T_{ij}\hat{\sigma}_{i}\otimes\hat{\sigma}_{j}],
\end{equation}
 where $\vec{M}$ and $\vec{N}$ are the Bloch vectors of the marginal states 
 and $T_{ij}=\text{Tr}( \hat{\sigma}_{i}\otimes\hat{\sigma}_{j}\rho_{AB})$ is the correlation function for the $i$th measurement of Alice and $j$th of Bob.  $T$ is the $3\times 3$ correlation matrix with entries $T_{ij}$. $T$ is invertible if its inverse matrix $T^{-1}$ exists. Using (\ref{EQ_2Q}), the figure of merit for RAC, i.e. the worst-case success probability, can be written as
\begin{eqnarray}
\label{n21minprob}
   p & = &  \min_{x\in\mathcal{X},y\in\mathcal{Y}} \frac{1}{4} \Big[\underbrace{\sum_{a,b}\delta_{g_{y}(a,b),x_{y}}}_{2\alpha_{xy}} +  \underbrace{\sum_{a,b}\delta_{g_{y}(a,b),x_{y}} (-1)^a}_{2\beta_{xy}} \hat{A}_{x}^{\dagger} \vec M  \\
   & + & \underbrace{\sum_{a,b}\delta_{g_{y}(a,b),x_{y}}(-1)^{b}}_{2\gamma_{xy}} \vec N^{\dagger}\hat{B}_{y} + \underbrace{\sum_{a,b}\delta_{g_{y}(a,b),x_{y}}(-1)^{a+b}}_{2\phi_{xy}} \hat{A}_{x}^{\dagger}T\hat{B}_{y} \Big], \nonumber
\end{eqnarray}
where the Bloch vectors $\hat{A}_{x}$, $\hat{B}_{y}$, $\vec M$ and $\vec N$ are now represented by $3\times 1$ column vectors, 
dagger denotes the transpose, and scalars $\alpha_{xy}$, $\beta_{xy}$, $\gamma_{xy}$, $\phi_{xy}$ are introduced to shorten the notation.
The quantum protocol outperforms the best classical one if $p>p_{\mathrm{cl}}$.
This inequality can be cast as conditions on components of encoding operations $\hat{A}_{x}$ in terms of decoding operations $\hat{B}_{y}$.
We explore solutions to this set of inequalities and in the following subsections prove the following results:
\begin{itemize}

\item  Sec.~\ref{secB} shows that a set of $2^{n}$ different encoding operations yielding quantum advantage for $n\geq 4$ does not exist.
\item Sec.~\ref{sec3} demonstrates that any two-qubit state with an invertible correlation matrix can be used to outperform the best classical protocol for $n=3$. We find explicitly encoding operations given noncoplanar Bloch vectors of decoding operations.

\item Sec.~\ref{secD} derives an additional (to the invertibility) condition that a correlation matrix has to satisfy to allow quantum codes which are more efficient than the best classical ones for $n=2$. We again provide encoding operations given any set of decoding operations.
\end{itemize}

\subsection{$n \ge 4$}
\label{secB}

It turns out that there are no useful RACs if $n \ge 4$. We first reduce the present RAC scenario to the case of RACs with quantum communication in order to utilise the result of Ref.~\cite{HI+06} and then provide an alternative proof.
By measuring her particle and communicating the classical bit Alice effectively prepares state $\rho_{B|a}$ in Bob's laboratory. Note that this state can be mixed, e.g. when shared state is not maximally entangled. 
The effect is as if Alice and Bob did not share a quantum state in advance and Alice just communicated a quantum particle in state $\rho_{B|a}$ to Bob. 
The final output of Bob is then determined by following a strategy described by a positive operator valued measure (POVM). 
Exactly this setting, i.e., communication of a mixed state on which Bob operates with a POVM, has been studied by Hayashi \emph{et al.} in \cite{HI+06} and it has been shown that such protocols do not admit $p > 1/2$ for $n \ge 4$. Below we provide an alternative proof. 
The proof of Hayashi \emph{et al.} uses geometric arguments whereas our method is algebraic.

\begin{result}
No $n\xrightarrow{p}1$ quantum RAC protocols with $n \geq 4$ admit $p > \frac{1}{2}$. 
\end{result}
\begin{proof}
We first derive a necessary condition on the (hypothetical) encoding operations $\hat{A}_{x}$ giving rise to $p > 1/2$ when a bipartite qubit state is shared between the parties.
Then we show that for $n \ge 4$ this condition is not satisfied. Recall that the worst-case success probability is given by
\begin{equation}
\label{n21minprobsuc}
p=\min_{x\in\mathcal{X},y\in\mathcal{Y}} \frac{1}{2}\Big[\alpha_{xy}+\beta_{xy}\hat{A}_{x}^{\dagger}\vec{M}+\gamma_{xy}\vec{N}^{\dagger}\hat{B}_{y}+\phi_{xy}\hat{A}_{x}^{\dagger}T\hat{B}_{y}\Big].
\end{equation}
The requirement $p > 1/2$ implies that for all pairs $(x,y)$ there exists a positive proper fraction $0<\chi_{xy}<1$ such that
\begin{equation}
\frac{1}{2}\Big[\alpha_{xy}+\beta_{xy}\hat{A}_{x}^{\dagger}\vec{M}+\gamma_{xy}\vec{N}^{\dagger}\hat{B}_{y}+\phi_{xy}\hat{A}_{x}^{\dagger}T\hat{B}_{y}\Big]=\frac{1}{2}\left(1+\chi_{xy}\right).
\end{equation} 
We rewrite this equation as 
\begin{equation}
\label{pairn21prob}
\hat{A}_{x}^{\dagger}\underbrace{\frac{[\beta_{xy}\vec{M}+\phi_{xy}T\hat{B}_{y}]}{||[\beta_{xy}\vec{M}+\phi_{xy}T\hat{B}_{y}]||}}_{\hat{B}_{xy}}=\underbrace{\frac{1+\chi_{xy}-\alpha_{xy}-\gamma_{xy}\vec{N}^{\dagger}\hat{B}_{y}}{||[\beta_{xy}\vec{M}+\phi_{xy}T\hat{B}_{y}]||}}_{\xi_{xy}}.
\end{equation}
where $||.||$ is the norm, $\hat{B}_{xy}$
 is normalized $3\times 1$ column vector and scalar $\xi_{xy}$ is introduced for future convenience.
 We now form a set of linear equations (\ref{pairn21prob}), each for a different value of $y$, and combine them into a matrix form
\begin{equation}
\label{n21encodingforeachx}
\hat{A}_{x}^{\dagger}B_{x}=\xi_{x},
\end{equation} 
where $B_{x}=[\hat{B}_{x1},\hat{B}_{x2},\hat{B}_{x3},\ldots,\hat{B}_{xn}]$ is a $3\times n$ matrix and $\xi_{x}=[\xi_{x1},\xi_{x2},\xi_{x3},\ldots,\xi_{xn}]$ is $1\times n$ row vector.
This set admits a formal solution
\begin{equation}
\label{midsolnAx}
\hat{A}_{x}=\Big(B_{x}B_{x}^{\dagger}\Big)^{-1}B_{x}\xi_{x}^{\dagger}.
\end{equation}
In Appendix \ref{app1} we show that for any $m\times n$ matrix $Q$ we have $(QQ^\dagger)^{-1}Q=Q(Q^{\dagger}Q)^{-1}$. Therefore, we can rewrite (\ref{midsolnAx}) as
\begin{equation}
\label{solnAx}
\hat{A}_{x}=B_{x}\Big(B_{x}^{\dagger}B_{x}\Big)^{-1}\xi_{x}^{\dagger}.
\end{equation}
Finally, we argue that for $n \ge 4$ this equation does not have a solution due to non-invertibility of $B_{x}^{\dagger}B_{x}$.

Since the columns of $B_{x}$ are normalized $3\times 1$ vectors the matrix $B_{x}^{\dagger}B_{x}$ is a Gram matrix with diagonal entries equal to one and off-diagonal entries being cosines of the angles between the Bloch vectors $\hat{B}_{xi}$ and $\hat{B}_{xj}$. It is known that a Gram matrix formed out of $n$ $m$-dimensional vectors is invertible if and only if the set of vectors is linearly independent~\cite{gramnote}. 
However, in the case of a qubit there are at most three linearly independent vectors on the Bloch sphere.  
Accordingly, $2^{n}$ distinct encodings of the form~(\ref{solnAx}) do not exist for $n\geq 4$. 
We can also use eigenvalue decomposition of $B_{x}^{\dagger}B_{x}$ to argue for the noninvertibility. Note that matrix $B_{x}^{\dagger}B_{x}$ is invertible if and only if all its eigenvalues are non-zero. 
However, in case of $B_{x}^{\dagger}B_{x}$ with $B_{x}$ being a $3\times n$ matrix, there are at most three non-zero eigenvalues, thus $B_{x}^{\dagger}B_{x}$ is non-invertible for $n\geq 4$. 
\end{proof}


The proof just given can also be extended to the cases where the encoding and decoding operations are two outcome POVMs. 
Since two outcome POVMs are convex combinations of projective measurements one still obtains Eqs.~(\ref{n21minprob}) and (\ref{n21minprobsuc}) but with different expressions for $\alpha_{xy},\beta_{xy},\gamma_{xy}$ and $\phi_{xy}$. 
Therefore, the same argument applies. 

Summing up, this subsection shows that the quantum codes studied here admit at most the same performance as the codes with communication of quantum particles.  It is not our aim to improve the worst-case probability of success of these quantum protocols. 
We rather wish to identify the properties of the shared quantum states that are responsible for beating the classical limits. 
In the next section we construct $3\xrightarrow{p}1$ and $2\xrightarrow{p}1$ RAC protocols demonstrating quantum advantage when a bipartite qubit state with invertible correlation matrix is shared between the parties. We focus on the following concrete decoding strategy: $g_{y}(a,b)= a \oplus b$ (sum modulo $2$). 
In this case $\alpha_{xy}=1$, $\beta_{xy}=\gamma_{xy}=0$ and $\phi_{xy}=(-1)^{x_{y}}$.
We begin with $3\xrightarrow{p}1$ RAC, where the sufficient condition for quantum advantage turns out to be very simple, and then move to $2 \xrightarrow{p}1$ case, where an additional constraint will be derived.

\subsection{$n = 3$}
\label{sec3}
  
The classical protocols give rise to $p_{\mathrm{cl}} = 1/2$ in this case and accordingly we look for the quantum protocols with the property:
\begin{equation}
     \label{opt31}
   p =  \min_{x\in\mathcal{X},y\in\mathcal{Y}}\left\lbrace\frac{1}{2}\left(1+(-1)^{x_{y}}\hat{A}_{x}^{\dagger}T\hat{B}_{y}\right)\right\rbrace > \frac{1}{2},
\end{equation}
where we have used Eq.~(\ref{n21minprob}) with $g_{y}(a,b)= a \oplus b$.
Equivalently, one needs to design strategies such that
   \begin{equation}
   \label{opt4}
(-1)^{x_{y}}\hat{A}_{x}^{\dagger}T\hat{B}_{y} > 0
   \end{equation}
   for each $x\in\{1,2,\ldots,8\}$ and $y\in\{1,2,3\}$.
They are characterised in the following theorem.
\begin{result}
   \label{remark2}
If a shared two-qubit state has invertible correlation matrix, then any $3\xrightarrow{p}1$ quantum RAC protocol with non-coplanar decoding Bloch vectors of Bob admits quantum advantage, i.e. $p>\frac{1}{2}$.
\end{result}
   \begin{proof}
We give an explicit construction of the encoding Bloch vectors of Alice.
 Namely, given a set of three decoding strategies $\mathcal{B}_{3} = \lbrace\hat{B}_{1},\hat{B}_{2},\hat{B}_{3}\rbrace$, 
 encoding corresponding to input $x$ is given by: 
 \begin{equation}
 \label{enc31}
          \hat{A}_{x}=\frac{(T^{-1})^{\dagger}[(-1)^{x_{1}}\hat{B}_{2}\times \hat{B}_{3}+ (-1)^{x_{2}}\hat{B}_{3}\times \hat{B}_{1}+ (-1)^{x_{3}}\hat{B}_{1}\times \hat{B}_{2}]}{||(T^{-1})^{\dagger}[(-1)^{x_{1}}\hat{B}_{2}\times \hat{B}_{3}+ (-1)^{x_{2}}\hat{B}_{3}\times \hat{B}_{1}+ (-1)^{x_{3}}\hat{B}_{1}\times \hat{B}_{2}]||},
 \end{equation}
 where we explicitly make use of the inverse of the correlation matrix, $T^{-1}$.
 For such strategies, taking $V$ as the volume of the parallelepiped formed using vectors $\mathcal{B}_{3}$ we have
\small
\begin{equation}
(-1)^{x_{y}}\hat{A}_{x}^{\dagger}T\hat{B}_{y}=\frac{V}{||T^{-1}[(-1)^{x_{1}}\hat{B}_{2}\times \hat{B}_{3}+ (-1)^{x_{2}}\hat{B}_{3}\times \hat{B}_{1}+ (-1)^{x_{3}}\hat{B}_{1}\times \hat{B}_{2}]||},
\end{equation}
\normalsize
which is indeed strictly positive and consequently $p>\frac{1}{2}$.
\end{proof}

As an illustration of this result consider bipartite qubit state in Eq.~(\ref{EQ_2Q}) with diagonal correlation matrix  $T=\text{diag}[t_{1},t_{2},t_{3}]$ and possibly non-zero local Bloch vectors.  Bob's decoding operations are taken to be along the Cartesian axes that diagonalize $T$. If $\{\hat{e}_{1},\hat{e}_{2},\hat{e}_{3}\}$ are decoding operations we have $T\hat{e}_{i}=t_{i}\hat{e}_{i}$ and for the inverse matrix, $T^{-1}$ we have $T^{-1}\hat{e}_{i}=\frac{1}{t_{i}}\hat{e}_{i}$. Using this decoding strategy Eq.~(\ref{enc31}) returns the following encoding strategy:
  \begin{equation}
  \label{eq:enc31}
  \hat{A}_{x} = \frac{1}{\sqrt{t_{1}^{-2}+t_{2}^{-2}+t_{3}^{-2}}}\left[\frac{(-1)^{x_{1}}}{t_{1}}\hat{e}_{1}+\frac{(-1)^{x_{2}}}{t_{2}}\hat{e}_{2}+\frac{(-1)^{x_{3}}}{t_{3}}\hat{e}_{3}\right],
  \end{equation}
   In this case the minimum success probability is given by
  \begin{equation}
  \label{qr14.1}
  p=\frac{1}{2}\left(1+\frac{1}{\sqrt{t_{1}^{-2}+t_{2}^{-2}+t_{3}^{-2}}}\right).
  \end{equation}
This probability has also been derived in Refs.~\cite{BP14,JD+19} for Bell diagonal states, i.e. states with vanishing local Bloch vectors.
As just shown it turns out that the formula generalises to any state.
This generalisation is possible because the function $g_y(a,b)$ we employed here leads to independence of the success probability on local Bloch vectors.
Furthermore, many of these states are not entangled.
Finally, we emphasise the simplicity of the condition found --- any state with invertible correlation matrix is useful in outperforming the classical solutions.

\subsection{$n = 2$}
\label{secD}

Note that the performance of classical $2\xrightarrow{p}1$ codes is better than the performance of classical $3\xrightarrow{p}1$ codes,
namely the worst-case success probability in the former protocols is $p_{\mathrm{cl}} = 2/3$ as compared to $1/2$ in the latter case.
We show here that for this reason the invertibility of the correlation matrix alone is not sufficient for quantum advantage when $n=2$. 
An additional condition will be derived. 
In particular, we will obtain the conditions that quantum states with invertible correlation matrix should satisfy to achieve $p> \frac{2}{3}$, given any set of decoding operations $(\hat{B}_{1},\hat{B}_{2})$. 
We denote the angle between them $\alpha$, i.e. $\hat{B}_{1}^{\dagger}\hat{B}_{2}=\cos\alpha$. 
As before, we proceed via explicit construction of encoding strategy for Alice characterised in the following theorem.
\begin{result}
\label{remark1}
If a shared two-qubit state has invertible correlation matrix and the square of its smallest singular value is greater than $\frac{1}{9}$,
then there exists a $2\xrightarrow{p}1$ quantum RAC protocol admitting quantum advantage, i.e., $p > \frac{2}{3}$.
\end{result}
\begin{proof}
Choose the encoding vectors as follows:
\begin{equation}
   \label{qr14.50}
   \hat{A}_{x} = \frac{(T^{-1})^{\dagger}[(-1)^{x_{1}}\hat{B}_{1}+(-1)^{x_{2}}\hat{B}_{2}]}{||(T^{-1})^{\dagger}[(-1)^{x_{1}}\hat{B}_{1}+(-1)^{x_{2}}\hat{B}_{2}]||}.
\end{equation}
Using eq.~(\ref{qr14.50}) the minimum success probability reads:
\begin{equation}
  \label{qr15}
  p = \frac{1}{2}\left[1+\min\left(\frac{\cos\frac{\alpha}{2}}{||(T^{-1})^{\dagger}\hat{B}_{+}||},\frac{\sin\frac{\alpha}{2}}{||(T^{-1})^{\dagger}\hat{B}_{-}||}\right)\right],
\end{equation}
where $\hat{B}_{\pm}=\frac{\hat{B}_{1}\pm\hat{B}_{2}}{||\hat{B}_{1}\pm\hat{B}_{2}||}$.
Let us arrange the singular values of $T$, given by $[t_{1},t_{2},t_{3}]$, in decreasing order $1\geq|t_{1}|\geq |t_{2}|\geq |t_{3}|$ and let $\hat{e}_{i}$ denote the eigenvector corresponding to the value $t_{i}$. 
We have $T\hat{e}_{i}=t_{i}\hat{e}_{i}$ and for the inverse matrix, $T^{-1}$ we have $T^{-1}\hat{e}_{i}=\frac{1}{t_{i}}\hat{e}_{i}$.  
Since $\lbrace \hat{e}_{1},\hat{e}_{2},\hat{e}_{3}\rbrace$ form a complete basis we expand $\hat{B}_{\pm}$ as follows
  \begin{align}
  \label{enc21}
  \hat{B}_{+} & = \cos\theta\hat{e}_{1}+\sin\theta(\sin\phi\hat{e}_{2}+\cos\phi\hat{e}_{3}),\\
  \label{enc22}
  \hat{B}_{-} & = -\sin\theta\hat{e}_{1}+\cos\theta(\sin\phi\hat{e}_{2}+\cos\phi\hat{e}_{3}),
  \end{align}
  and use this parameterisation in the figure of merit
  \begin{equation}
  \label{qr16}
  p=\frac{1}{2}\left[1+\min\left(\frac{\cos\frac{\alpha}{2}}{\sqrt{K_{+}}},\frac{\sin\frac{\alpha}{2}}{\sqrt{K_{-}}}\right)\right],
  \end{equation} 
  where
  \begin{subequations}
  \begin{eqnarray}
  \label{qr17}
  K_{+} & = & \frac{\cos^{2}\theta}{t_{1}^{2}}+\frac{\sin^{2}\theta\sin^{2}\phi}{t_{2}^{2}}+\frac{\sin^{2}\theta\cos^{2}\phi}{t_{3}^{2}}, \\
  \label{qr18}
  K_{-} & = & \frac{\sin^{2}\theta}{t_{1}^{2}}+\frac{\cos^{2}\theta\sin^{2}\phi}{t_{2}^{2}}+\frac{\cos^{2}\theta\cos^{2}\phi}{t_{3}^{2}}.
  \end{eqnarray}
  \end{subequations}
  To beat the classical performance $\min\left(\frac{\cos\frac{\alpha}{2}}{\sqrt{K_{+}}},\frac{\sin\frac{\alpha}{2}}{\sqrt{K_{-}}}\right) > 1/3$, i.e. each argument must satisfy
\begin{equation}
\label{k+}
\frac{\cos\frac{\alpha}{2}}{\sqrt{K_{+}}}>\frac{1}{3}, \quad
\frac{\sin\frac{\alpha}{2}}{\sqrt{K_{-}}}>\frac{1}{3}.
\end{equation}
From these inequalities we obtain 
\begin{equation}
\label{2to1quantadv}
\frac{1}{9}K_{+}<\cos^{2}\frac{\alpha}{2}<1-\frac{1}{9}K_{-}.
\end{equation}
Since $K_{\pm}$ are the convex combinations of $\frac{1}{t_{1}^{2}},\frac{1}{t_{2}^{2}},\frac{1}{t_{3}^{2}}$ and
for any $\xi$ which is a convex combination of $\lbrace \xi_{1}, \xi_{2},\dots,\xi_{n}\rbrace$, we have $\xi \leq \max\lbrace \xi_{1}, \xi_{2},\dots, \xi_{n}\rbrace$,
it follows that $K_{\pm} \le 1/ t_3^2$.
Therefore, for states with $t_{3}^2 > 1/9$ the lower bound in Eq.~(\ref{2to1quantadv}) is less than $1$ and the upper bound is more than $0$. 
In such cases there exists angles $\alpha$ giving rise to the quantum advantage, i.e. $p> \frac{2}{3}$.
\end{proof}
Note that we have compared the quantum RAC to the corresponding classical RAC assisted with two classical bits in an arbitrary distribution, i.e. the case (ii) in Sec.~\ref{cv} with the worst case success probability $p_{\mathrm{cl}} = 2/3$. The quantum solution discussed here does not depend on the Bloch vectors of the marginal states and therefore the conditions derived hold as well for states with local maximally mixed states i.e, Bell diagonal states. In this case it is fair to compare their performance to the classical codes assisted with bits having maximally mixed marginals. These have been described as case (iii) in Sec.~\ref{cv} and admit $p_{\mathrm{cl}} = 1/2$, just as the codes for $n = 3$. In such a case the invertibility of correlation matrix alone is the sufficient condition for quantum advantage.\\

As an example of how separable quantum state can outperform the best classical strategy we consider the generalization of the $2\xrightarrow{p}1$ code discussed in Refs.~\cite{PZ10,BP14}. 
Let the shared state be Bell-diagonal i.e., $\rho_{AB}=\frac{1}{4}[\mathbb{\hat{I}}\otimes\mathbb{\hat{I}}+\sum_{i}t_{i}\hat{\sigma}_{i}\otimes\hat{\sigma}_{i}]$ with $-1\leq t_{i}\leq +1$ and $|t_{1}|\geq |t_{2}|\geq |t_{3}|$. This state is separable if $|t_{1}|+|t_{2}|+|t_{3}| \leq 1$~\cite{Horodecki1996}.  Let the decoding strategies be orthogonal i.e., $\alpha=\frac{\pi}{2}$. 
From (\ref{qr16}) the worst-case success probability is
\begin{equation}
\label{2to1bd}
p=\frac{1}{2}\left[1+\frac{1}{\sqrt{2}}\min\left(\frac{1}{\sqrt{K_{+}}},\frac{1}{\sqrt{K_{-}}}\right)\right],
\end{equation}
where $K_{\pm}$ is given by  (\ref{qr17}) and (\ref{qr18}).  
The codes of Ref.~\cite{PZ10,BP14} are recovered if we put $\theta=\frac{\pi}{4}$ and $\phi=\frac{\pi}{2}$ in the expressions for $K_{\pm}$. 
Since we are considering quantum states with locally maximally mixed marginals, the condition for quantum advantage is $p>\frac{1}{2}$, or equivalently
\begin{equation}
\label{2to1bdmin}
\min\left(\frac{1}{\sqrt{K_{+}}},\frac{1}{\sqrt{K_{-}}}\right) >  0.
\end{equation}
For Werner states, where $t_{1}=t_{2}=t_{3}=-k$ and $0\leq k\leq 1$, we have $K_{\pm}=|k|$ and the minimum success probability is $p= \frac{1}{2}(1+\frac{|k|}{\sqrt{2}})$ which is greater than half for all non-zero $k$. At the same time the state is separable for $k \le 1/3$, but admits quantum discord in the whole range of $k$.


 
 \section{Discussion}
 
As shown, there exist non-zero discord Bell diagonal states that allow for the creation of efficient quantum codes for $n=2,3$. This was also found in Ref.~\cite{BP14}. Building on that, it was pointed out in Ref.~\cite{JD+19} that the amount of quantum discord does not specify the Bell diagonal state yielding optimal efficiency for the RAC protocol discussed in \cite{PZ10,BP14}. For this reason the authors constructed a suitable measure of non-classicality whose maximum value yields the most efficient Bell diagonal quantum states~\cite{JD+19}.  However, none of these works went beyond Bell diagonal states. In particular, it is unclear whether quantum discord is sufficient to guarantee existence of efficient $n\xrightarrow{p}1$ quantum codes for $n=2,3$. In this regards, the present work proposes a general sufficient criteria for quantum advantage applicable to any bipartite qubit state. 
We find that in the case of $3\xrightarrow{p}1$ RACs any assisting state with invertible correlation matrix can be exploited to outperform the best classical protocol.
For $2\xrightarrow{p}1$ codes (due to better performance of the classical codes) the invertibility has to be augmented with a suitable condition on the smallest singular value of the correlation matrix.
Recall that a matrix is invertible if and only if all the eigenvalues are non-zero. Thus, if the bipartite qubit state has vanishing quantum discord, the correlation matrix is non-invertible, but the converse statement does not always hold~\cite{LM+10}. 
For example, geometric discord of Bell diagonal states equals $D_G = \frac{1}{4}(t_2^2 + t_3^2)$ in our present notation~\cite{Dakic2010}.
Therefore states with $t_2 \ne 0$ but $t_3 = 0$ provide examples of discorded states with non-invertible $T$ matrix.
These states do not satisfy our sufficient condition for efficient RACs despite being discorded and hence they are natural starting point for further study.
We also note that in the case of the bipartite qubit state, the maximum amount of mutual information that can exist simultaneously among all the members of a triad of mutually unbiased bases is non-vanishing if and only if the correlation matrix is invertible \cite{GW14}. Therefore, the invertibility of the correlation matrix is a non-classical feature of correlations in a quantum state.

In conclusion, we studied resources behind quantum advantage in $n\xrightarrow{p}1$ RAC protocols assisted with arbitrary two-qubits states. For $n \ge 4$ there are no useful protocols, independently of the assisting state. For smaller codes, with $n = 2$ and $n = 3$, we demonstrated explicitly that advantage exists and that invertibility of correlation matrix is the key property behind quantum performance. The invertibility has also been linked to various forms of quantum correlations. These results provide a step towards the possibility of using invertibility of correlation matrix as resource for a quantum information theoretic tasks. 
 
 \section*{Acknowledgments}
 SK acknowledges NASI Research Associate fellowship and QUEST-DST fellowship for their support. CJ acknowledges S. N. Bose Centre, Kolkata for the postdoctoral fellowship. The research of DH is supported by NASI Senior Scientist fellowship and the QUEST-DST project of the Govt. of India. 
\appendix

\section{Proof of the classical limits for $n\xrightarrow{p}1$ Random Access Codes when Alice and Bob shares classical bits from a source}
\label{app0}

Here we briefly recall the proof of the classical performance of $n\xrightarrow{p}1$ RACs when Alice and Bob shares classical bits from a source, as given in \cite{BP14}. 
Consider the scenario where Alice and Bob receive bits $r_{a}$ and $r_{b}$ respectively and Alice has bit-string $x=[x_{1},x_{2},\ldots,x_{n}]$ from a set of $2^{n}$ $n$-bit strings $\mathcal{X}$. 
If Bob's guess for the $y$th bit is $b_{y}$, where $y\in\mathcal{Y}$, then the minimum success probability is given by
\begin{align}
\label{minclass}
p& = \min_{x\in\mathcal{X},y\in\mathcal{Y}}\sum_{r_{a},r_{b}}p_{r_{a}.r_{b}} \, \mathrm{Pr}(b_{y}=x_{y}|x,r_{a},r_{b})
\end{align}
where $p_{r_{a},r_{b}}$ is the probability that Alice and Bob receive bits $r_{a}$ and $r_{b}$ from the shared source.

Let Alice's classical communication be $c(x,r_{a})$, then Bob's guess for the $y$th bit is the function of $c$ and $r_{b}$ i.e. $b_{y}=b_{y}(c,r_{b})$. 
For any given input $x$ Alice can only choose from the four deterministic encoding strategies: $1)$ $c_x=0$ for all values of $r_{a}$, $2)$ $c_x=1$ for all values of $r_{a}$, $3)$ $c_x=r_{a}$ and $4)$ $c_x=1\oplus r_{a}$. 
Consider Alice chooses the same encoding for two different bit strings $x$ and $x'$, i.e. $c_x = c_{x'}$.
Then Bob cannot distinguish between $x$ and $x'$ from $c$. If the minimum success probability for string $x$ is greater than half then the probability for all the bits in $x$ must be greater than half. 
Consequently, the success probability for the bits in $x'$  which are different than those of $x$ must be less than half (for a fixed $r_b$ as well as averaged over the shared randomness). 
Since we assume $r_a$ is a binary variable, for $n \ge 3$ Alice is forced to assign the same encoding to different $x$ and $x'$, and hence $p = 1/2$.

Now we focus on $n=2$. 
We arrange Bob's guesses into a vector $(b_{1},b_{2})=g_{c,r_{b}}$.
For a bit-string $x$ let us define probability vector
\begin{equation}
\label{appprobvec}
P(x)=(\mathrm{Pr}(b_{1}=1|x),\mathrm{Pr}(b_{2}=1|x)),
\end{equation}
 where $\mathrm{Pr}(b_{y}=1|x)$ is the probability that Bob's guess $b_{y}$ is equal to one, i.e.  $\mathrm{Pr}(b_{y}=1|x)=\sum_{r_{a},r_{b}}p_{r_{a},r_{b}} \mathrm{Pr}(b_{y}=1|r_{a},r_{b},x)$. 
 Note that,  given $r_{a}$, $r_{b}$ and $x$, Bob's guess is deterministic
\begin{equation}
\label{appdeter}
\mathrm{Pr}(b_{y}=1|r_{a},r_{b},x)=b_{y}.
\end{equation} 
Consequently, we can re-write the probability vector as
\begin{align}
\label{detprobvec}
P(x)& = \sum_{r_{a},r_{b}}p_{r_{a},r_{b}}(b_{1},b_{2})\\
\label{detprobvec1}
& =\sum_{r_{a},r_{b}}p_{r_{a},r_{b}}g_{c,r_{b}}
\end{align}
If there does not exist any pair $(r_{a},r_{b})$ such that $g_{c,r_{b}}=x$ then Bob will always predict one of the bits wrongly. 
The probability vector for each of the encoding functions of Alice is given by
\begin{align}
\label{c0}
P_{1}(x) & =p_{0,0}g_{0,0}+p_{0,1}g_{0,1}+p_{1,0}g_{0,0}+p_{1,1}g_{0,1}\\
\label{c1}
P_{2}(x) & =p_{0,0}g_{1,0}+p_{0,1}g_{1,1}+p_{1,0}g_{1,0}+p_{1,1}g_{1,1}\\
\label{c2}
P_{3}(x) & =p_{0,0}g_{0,0}+p_{0,1}g_{0,1}+p_{1,0}g_{1,0}+p_{1,1}g_{1,1}\\
\label{c3}
P_{4}(x) & =p_{0,0}g_{1,0}+p_{0,1}g_{1,1}+p_{1,0}g_{0,0}+p_{1,1}g_{0,1}
\end{align}
As shown above, she needs to use different encoding for different strings $x$.
Let us denote the $j$th string by $x^j$ and consider probability vectors $P_j(x^j)$, i.e. the $j$th encoding is applied to the $j$th input string.
Since there are four input strings and four arguments $(c, r_b)$ the values of $g_{c,r_b}$ must be different for different pairs $(c,r_b)$ to match $g_{c,r_b} = x$.
Optimising the probabilities under these constraints one finds that it is possible to use the bias in the distribution of random bits $r_a$ and $r_b$ to completely avoid giving guesses that have both individual bits wrong.
The optimal worst case success probability is $p = 2/3$.

 \section{Proof of $(QQ^\dagger)^{-1}Q=Q(Q^{\dagger}Q)^{-1}$}

\label{app1}
\begin{thm}
Any $m\times n$ matrix $Q$ satisfies
\begin{equation}
\label{claim}
(QQ^\dagger)^{-1}Q=Q(Q^{\dagger}Q)^{-1}
\end{equation}
\end{thm}
\begin{proof}
Without loss of generality we assume $m\leq n$.

 First consider Q to be a $m\times n$ diagonal matrix i.e. $Q=\text{diag}[q_{ii}]$, where $1 \leq i \leq m$. As $Q$ is a diagonal matrix its transpose $Q^{\dagger}$ is $n\times m$ matrix with the same elements i.e. $Q^{\dagger}= \text{diag}[q_{ii}]$, where $1 \leq i \leq m$. Then $QQ^{\dagger}$ is a $m\times m$ diagonal matrix with the entries $q_{ii}^{2}$ i.e. $QQ^{\dagger}=\text{diag}[q_{ii}^{2}]$, where $1\leq  i \leq m$. The inverse of $QQ^{\dagger}$ is a $m\times m$ matrix with the entries $\frac{1}{q_{ii}^{2}}$, i.e $(QQ^{\dagger})^{-1}= \text{diag}[\frac{1}{q_{ii}^{2}}]$, where $1\leq  i \leq m$. $(QQ^{\dagger})^{-1}Q$ is a $m\times n$ diagonal matrix with the entries $\frac{1}{q_{ii}}$, i.e. $(QQ^{\dagger})^{-1}Q = \text{diag}[\frac{1}{q_{ii}}]$, where $1\leq  i \leq m$. Following the same process it can be shown that $(Q^{\dagger}Q)^{-1}$ is $n\times n$ diagonal matrix with non-zero diagonal elements for the first $m$ rows, i.e. $(Q^{\dagger}Q)^{-1}=\text{diag}[\frac{1}{q_{ii}^{2}}]$, where $1\leq  i \leq m$. Then it can be seen that $Q(Q^{\dagger}Q)^{-1}$ is $m\times n$ diagonal matrix with entries $\frac{1}{q_{ii}}$, i.e. $Q(Q^{\dagger}Q)^{-1}=\text{diag}[\frac{1}{q_{ii}}]$, where $1\leq  i \leq m$. Therefore,  $(QQ^\dagger)^{-1}Q=Q(Q^{\dagger}Q)^{-1}$ when $Q$ is a diagonal matrix.\\
 
 For a general $m\times n$ matrix $Q$ we use the singular value decomposition of $Q$ i.e. $Q=U\Sigma_{Q}V^{\dagger}$ where $\Sigma_{Q}$ is the $m\times n$ diagonal matrix, $U$ and $V$ are $m\times m$ and $n\times n$ unitary matrices respectively. The LHS of eq.~(\ref{claim}) can now be written as
 \begin{align}
 \label{lhs1}
 (QQ^\dagger)^{-1}Q & = (U\Sigma_{Q}\Sigma_{Q}^{\dagger}U^{\dagger})^{-1}U\Sigma_{Q}V^{\dagger}\\
 \label{lhs2}
 & = U(\Sigma_{Q}\Sigma_{Q}^{\dagger})^{-1}\Sigma_{Q}V^{\dagger}
 \end{align}
 Similarly, the RHS of eq.~(\ref{claim}) can be written as
  \begin{align}
  \label{rhs1}
  Q(Q^\dagger Q)^{-1} & = U\Sigma_{Q}V^{\dagger}(V\Sigma_{Q}^{\dagger}\Sigma_{Q}V^{\dagger})^{-1}\\
  \label{rhs2}
  & = U\Sigma_{Q}(\Sigma_{Q}^{\dagger}\Sigma_{Q})^{-1}V^{\dagger}
  \end{align}
Since $\Sigma_{Q}$ is a diagonal matrix we have already seen that $(\Sigma_{Q}\Sigma_{Q}^{\dagger})^{-1}\Sigma_{Q}=\Sigma_{Q}(\Sigma_{Q}^{\dagger}\Sigma_{Q})^{-1}$, thus (\ref{lhs2}) and (\ref{rhs2}) are equal.
\end{proof}

 \bibliography{cite}

 \end{document}